\documentclass[
    ,final            
  ]
  {aipproc}

\layoutstyle{6x9}

\begin{document}

\title{On Scalar Meson Dominance in Semi- and Non-Leptonic Weak Pseudoscalar Meson Decays}


\author{Frieder Kleefeld \footnote{e-mail: kleefeld@cfif.ist.utl.pt, URL: http://cfif.ist.utl.pt/$\sim$kleefeld/}}{
  address={Centro de F\'{\i}sica das Interac\c{c}\~{o}es Fundamentais (CFIF), Instituto Superior T\'{e}cnico,\\
Edif\'{\i}cio Ci\^{e}ncia, Piso 3, Av. Rovisco Pais, P-1049-001 LISBOA, Portugal}
} 

\begin{abstract}
Aspects of Scalar Meson Dominance in semi- and non-leptonic weak pseudoscalar meson decays are shortly discussed on the basis of the Quark-Level Linear Sigma Model.
\end{abstract}

\maketitle


{\bf Introduction.} The theoretical description of leptonic, semi- and non-leptonic weak decays of pseudoscalar mesons has been known to be extremly rewarding since at least 5 decades \cite{Buras:1992aj} in particular due to technical and numerical challenges in the description of strong interactions on the basis Quantum Chromodynamics (QCD) \cite{Verma:1996wq}. The experimental and theoretical study of non-leptonic $K$-meson decays has for instance a long tradition \cite{Buras:1992aj,Cabibbo:1964qq,Goswami:1972rn,Fayyazuddin:1974ud,Shifman:1976ge}, yet the theoretical issues concerning these decays are far from settled \cite{Scadron:1987vk,Pallante:2000hk,Cheng:1986an}. To use the words of G.\ Colangelo \cite{Cheng:1986an}: \emph{``$\ldots$ The calculation of the $K\rightarrow \pi\pi$  amplitude in the Standard Model still remains one of the most difficult and yet unsolved problems of today's particle physics, despite many years of efforts and progress in our understanding of various related physics aspects. Indeed we lack yet a satisfactory explanation of the $\Delta I=1/2$ rule, and do not yet have a calculation  of $\varepsilon^\prime/\varepsilon$ in the Standard Model which would make a comparison to the measured value somewhat useful --- the typical size of the theoretical uncertainties attached to any of the calculations available in the literature is around 50 \% $\ldots$.  A recurring disussion in the literature on calculations of $K\rightarrow \pi\pi$  concerns the role of final state interactions (FSI) $\ldots$''}. 
In view of the outlined situation we want to recall that a growing amount of meson production data available from charm/bottom quark factories allows for the first time a direct view \cite{Bugg:2005xx} to the nature and properties of light scalar mesons which up now were only hardly accessable in elastic meson-meson scattering due to the unfortunate presence of Adler-zeros \cite{Rupp:2004rf} and were taken into account in QCD-based formalisms to meson weak decays only indirectly as hardly quantifiable non-perturbative effects. The very existence of light scalar mesons draws renewed attention to the strongly neglected \emph{Partial} Conservation of flavour-changing Vector Current (PCVC) which has coexisted with often celebrated Partial Conservation of Axialvector Current (PCAC) for about 4 decades \cite{Gerstein:1969fc}. We shall here report shortly on first results of a ``Scalar Meson Dominance'' (SMD) approach to pseudoscalar meson weak decays taking into account PCVC manifestly by including scalar mesons, rather than guessing PCVC non-perturbatively like in QCD. 

{\bf The Linear Sigma Model (L$\sigma$M) and its dynamical generation.}
Facing \cite{Kleefeld:2005hf,Kleefeld:2002au} the yet lacking conclusive experimental evidence for the existence of gluons, new developments in mathematical physics and puzzling experimental data we will follow here Ref.\ \cite{Kleefeld:2005hd} and model strong interactions on the basis of an alternative approach to QCD known under the name ``Quark-Level Linear Sigma Model'' (QLL$\sigma$M) \cite{Levy:1967a}, which makes use of degrees of freedom being found in nature, i.e.\  mesons and (anti)quarks. In other words we dynamically generate as in Ref.\ \cite{Kleefeld:2005hd} our strong interaction Lagrangean on the basis of the following Yukawa-like interaction Lagrangean relating (anti)quark fields (denoted by $q_\pm(z)$) and scalar ($S(z)$), pseudoscalar ($P(z)$), vector ($V(z)$), and axialvector ($Y(z)$) $U(6)\times U(6)$ meson field matrices in flavour space (see also \cite{Kleefeld:2002au}) (The undetermined signs $s_s$, $s_p$, $s_v$, $s_y\in \{-1,+1\}$ are here irrelevant!):
\begin{eqnarray} \lefteqn{{\cal L}^{\,strong}_{\,int} (z) =} \nonumber \\
 & = &  \sqrt{2}\, g \;\, \overline{q^c_+} (z)  \left( s_s \, S(z) + s_p \,i\,P(z)\gamma_5 + \frac{e^{-i\,\alpha}}{2}  \Big(s_v \not\!{V}(z) + s_y \not\!{Y}(z)\, \gamma_5\Big) \right)  q_-(z) \; , \label{eqlagrstrong1}
\end{eqnarray}
with $g=|g|\exp(i\alpha)$ being the eventually complex strong interaction coupling constant. The dynamical generation of an effective meson-meson interaction Lagrangean (like e.g.\ the well known $U(3)\times U(3)$ L$\sigma$M \cite{Levy:1967a,Kleefeld:2005qs}) makes then use --- to avoid double counting --- of the idea that meson-meson interaction proceeds through quark-loops only and {\em not on the basis of any further extra direct meson-meson interaction terms} in the Lagrangean.
${\cal L}^{\,strong}_{\,int} (z)$ is now added --- as done by Cabibbo \& Maiani \cite{Levy:1967a} --- to the Lagrangean describing (in Fermi's limit) weak interactions among leptons (denoted here as $\ell_\pm (x)$) and (anti)quarks, i.e. ($G_F=$ Fermi's constant, $\theta_W=$ Weinberg's angle, $T_3=$ isospin matrix, $Q_q, Q_\ell=$ charge matrices, $V_{{}_{CKM}}=$ CKM-matrix, $P_{{}_L}\equiv(1-\gamma_5)/2$):
\begin{eqnarray} \lefteqn{{\cal L}_{\,int}^{\,weak} (x) \; = \;  - \; 4\, \frac{G_F}{\sqrt{2}} \; \times} \nonumber \\
 & \times & \Big[\,\Big( \, \overline{\ell^c_+}\, (x) \; \gamma^\mu \; P_{{}_L} \left( \begin{array}{cc} 0_3 & 1_3 \\ 0_3 & 0_3 \end{array} \right) \ell_- (x) \; + \;
\overline{q^c_+}\,(x) \; \gamma^\mu \;
P_{{}_L}  \left( \begin{array}{cc} 0_3 & V_{{}_{CKM}} \\ 0_3 & 0_3 \end{array} \right) q_-(x) \Big) \nonumber \\
 & & \;\;\,  \,\Big( \overline{\ell^{\,c}_+}\, (x) \; \gamma_\mu \; P_{{}_L} \left( \begin{array}{cc} 0_3 & 0_3 \\ 1_3 & 0_3 \end{array}\right) \ell_-(x) \; + \;
 \overline{q^c_+}\,(x) \; \gamma_\mu \;
P_{{}_L} \left( \begin{array}{cc} 0_3 & 0_3 \\ \overline{V}_{{}_{CKM}} & 0_3 \end{array}\right) q_-(x) \Big) \nonumber \\
 & & +  \Big( \overline{\ell^c_+}\,(x) \; \gamma^\mu ( T_3 \, P_{{}_L}  - Q_\ell\, \sin^2 \theta_W ) \, \ell_-(x) +  
\overline{q^c_+}\,(x) \; \gamma^\mu ( T_3 \, P_{{}_L}  - Q_q\, \sin^2 \theta_W ) \; q_-(x) \Big)  \nonumber \\
 & & \;\;\;\, \Big( \overline{\ell^c_+}\,(x) \; \gamma_\mu ( T_3 \, P_{{}_L}  - Q_\ell\, \sin^2 \theta_W ) \; \ell_-(x) +  
\overline{q^c_+}\,(x) \; \gamma_\mu ( T_3 \, P_{{}_L}  - Q_q\, \sin^2 \theta_W ) \; q_-(x) \Big) \Big] \nonumber \\[1mm]
 & + & \mbox{h.c.} \; .
\end{eqnarray}

{\bf Leptonic weak meson decays and pseudoscalar decay constants.}
As a first step we study leptonic decays of pseudoscalar mesons to extract pseudoscalar decay constants $f_P$. The relevant part $S_{e\!f\!f}[P\,\bar{\ell}\ell]$ of the effective action in the local limit resulting by dynamical generation is given in Ref.\ \cite{Kleefeld:2005hd}. By inspection of this effective action we conclude as in Ref.\ \cite{Kleefeld:2005hd} that the finite decay constant $f_{\eta_{q_1\bar{q}_2}}$ of a pseudoscalar meson $\eta_{q_1\bar{q}_2}$ corresponds in the considered formalism to a log.-divergent integral, i.e.\ \cite{Kleefeld:2005hd}\footnote{The colour factor $N_c$ is displayed here for traditional reasons and can absorbed by rescaling $g$.} \footnote{This replacement of logarithmic divergences by finite experimental numbers renormalizes the formalism without the need of introducing unpleasant regulators like cutoffs.}:
\begin{equation} f_{\eta_{q_1\bar{q}_2}}  \longleftrightarrow \; - \, 4 \, i\,  N_c \, |g| \int \frac{d^4p}{(2\pi)^4} \;  \frac{(m_{q_1} + m_{\bar{q}_2})/2}{(p^2- m_{q_1}^2)(p^2- m_{\bar{q}_2}^2)} \; . \label{pseudconsteq1}
\end{equation}

{\bf Semi-leptonic weak meson decays and SMD.} 
As a first instructive example we consider in Ref.\ \cite{Kleefeld:2005hd} the semi-leptonic decay $K^+\rightarrow \pi^0\,e^+\nu_e$. Ingredient for our calculation has been the dynamical generation of the effective action $S_{e\!f\!f}[S\,PP]$ between one scalar and two pseudoscalar mesons displayed in the footnote \footnote{In the considered local limit we obtained \cite{Kleefeld:2005hd} ($M_q\equiv\mbox{diag}[m_u,m_c,m_t,m_d,m_s,m_b]$, ``$\mbox{tr}_{{}_F}$''= flavour trace):
\begin{eqnarray} 
\lefteqn{S_{e\!f\!f}[S\,PP] \; = \; \int d^4z \;\; \sqrt{2} \;\;  g^2 \;\,e^{i\,\alpha} \,\; s_s \,  \; (- 4\; i \; N_c \; |g|) \; \int \frac{d^4p}{(2\pi)^4} } \nonumber \\
 & \times & \Big\{ \mbox{tr}_F \Big[ \, S(z) \; \frac{1}{(p^2-M^2_q)}\; \{\, P^2(z)\,,\,M_q\, \} \; \frac{1}{(p^2-M^2_q)} \; \Big] \nonumber \\
 & & + \mbox{tr}_F \Big[ \, [\,S(z) \, , \, P(z)\,] \; \frac{1}{(p^2-M^2_q)}\; [\, P(z)\,,\,M_q\, ] \; \frac{1}{(p^2-M^2_q)} \; \Big] \nonumber \\
 & &  - \,\mbox{tr}_F \Big[ \,  \{ \, S(z)\, , \, M_q \, \} \; \frac{1}{(p^2- M_q^2)} [\, P(z)\,,\,M_q\, ] \; \frac{1}{(p^2-M^2_q)}\; [\, P(z)\,,\,M_q\, ] \; \frac{1}{(p^2-M^2_q)} \; \Big] \Big\}  +  \ldots \, .
\end{eqnarray}.}. It is then straight forward to dynamically generate the (local) effective action for the process $K^+\rightarrow \pi^0\,e^+\nu_e$ \cite{Kleefeld:2005hd}:
\begin{eqnarray} S_{e\!f\!f}& =& \int d^4z \;\; (-i\,e^{2\, i\,\alpha}) \; \left( - \frac{G_F}{\sqrt{2}} \right) \; \overline{V}_{us}\;\; \overline{e^{\,c}_+}\, (z) \; \gamma_\mu \, (1-\gamma_5) \; \nu_{e\,-}(z) \nonumber \\
 & \times &  \frac{1}{\sqrt{2}} \;\Big\{ \, \pi^0(z) \, \left( \, \frac{2\,|\,g|\,f_{K^+}}{m_u + m_s} \;\, \partial^\mu \, K^+(z)\right) \, - \, K^+(z) \, \left( \, \frac{2\,|\,g|\,f_{\eta_{u\bar{u}}}}{m_u + m_u} \;\, \partial^\mu \,\pi^0(z)\right) \nonumber \\
 & & + 4 i \, N_c \, |\,g|^2 \, (m_s - m_u)^2 \; K^+(z) \, ( \partial^\mu \, \pi^0(z)) \int \frac{d^4p}{(2\,\pi)^4}\, \frac{1}{(p^2-m^2_s)(p^2-m^2_u)^2} \nonumber \\
 & & + \frac{\lambda}{g^2} \, m_s \, \frac{(m_s - m_u)}{m^2_{\kappa^+}} \, \frac{2\,|\,g|\,f_{K^+}}{m_u + m_s} \, \Big( \pi^0(z) \, (\partial^\mu \, K^+(z))  +  K^+(z) \, (\partial^\mu \, \pi^0(z) )   \Big)   \Big\} \nonumber \\
 & + & \mbox{$K^\ast$-exchange} + \ldots \, .
\end{eqnarray}
For $\lambda\simeq 2 g^2$ the weak transition formfactors to relevant order in the scale $\delta = (m_s/m_u) - 1\simeq 0.44\;$  according to the nonrenormalization theorem of Ademollo \& Gatto \cite{Ademollo:1964sr} are determined as \cite{Kleefeld:2005hd} $f^{\,K^+\pi^0}_+(0)= 1 + O(\delta^2)$ and
$f^{\,K^+\pi^0}_-(0) = 4\,|g| \,e^{2 i \alpha} \, \delta \, (1+\delta)\; (2+\delta)^{-1}  |m_u||f_{K^+}| m^{-2}_{\kappa^+} + O(\delta^2)$ displaying SMD. Because of \cite{Eidelman:2004wy} $f^{\,K^+\pi^0}_-(0)/f^{\,K^+\pi^0}_+(0) \simeq - 0.125 \pm 0.023$, $|f_{K^+}|\simeq 159\;\mbox{MeV}/\sqrt{2}$, $m_{\kappa^+} \simeq 797$ MeV, and $|m_u|\simeq 337$ MeV there has to hold $0> e^{2 i \alpha}\simeq -1$ due to the 1-loop dynamically generated value $|g|=2\pi/\sqrt{3}$ \cite{Kleefeld:2005hd}. In conclusion experiment is suggesting \cite{Kleefeld:2005hd} an \emph{asymptotic free} QLL$\sigma$M with Yukawa coupling $g\simeq -i\,2\pi/\sqrt{3}$ and quartic coupling $\lambda\simeq 2\,g^2 \simeq - 8\pi^2/3<0$, respectively.  

{\bf Non-leptonic weak meson decays and SMD.} For the theoretical explanation of two longstanding problems based on non-leptonic meson weak decays, i.e.\ the quantitative calculation of $|M(K^+\rightarrow \pi^+\pi^0)|$ and the so-called $\Delta I=1/2$ rule, there might exist an attractive solution on the basis of SMD \footnote{The observation that the ``vacuum saturation'' (VS) W-emission contribution to the non-leptonic decay $K^+\rightarrow \pi^+\pi^0$ predicts an amplitude \emph{two times} larger than the experimental value $|M_{K^+\rightarrow \pi^+\pi^0}(exp)|  = m_{K^+} \, (8\pi\,\Gamma/p_{cm})^{1/2} \; = \; (1.832 \pm 0.007)\cdot 10^{-8} \;\mbox{GeV}$ led SVZ \cite{Shifman:1976ge} to decrease the discrepancy between theory and experiment on the basis of so-called ``penguin-diagrams'' \cite{Shifman:1976ge} which are now constraint strongly by the fact that \emph{``polychromatic} [i.e.\ QCD] \emph{penguins don't fly''} \cite{Chivukula:1986du}. 1987 Scadron proposed a VS being \emph{half} of the VS of SVZ and agreeing to experiment, as $|M_{K^+\rightarrow \pi^+\pi^0}(Scadron)|=|G_F V_{ud} \overline{V}_{us} ( f_{\pi^-} \, (m^2_{K^+} - m^2_{\pi^0}) - 2\, f_{K^+} \, (m^2_{\pi^+} - m^2_{\pi^0}))/(2\sqrt{2})|=(1.886 \ldots\, -\,0.026\ldots)\cdot 10^{-8} \;\mbox{GeV}=1.860 \ldots \cdot 10^{-8} \;\mbox{GeV}$. Scadron's mysterious factor 1/2 might be explained by adding to the SVZ W-emission contribution SMD meson-exchange diagrams like in Fig.\ 1 of Ref.\ \cite{Goswami:1972rn}. Unfortunately these extra contributions require a yet unexplained mechanism to mix scalar and pseudoscalar mesons ($S$-$P$ mixing). Another testing ground for SMD in non-leptonic meson weak decays is the so-called ``$\Delta I=1/2$ rule'' \cite{Buras:1992aj,Fayyazuddin:1974ud,Shifman:1976ge,Cheng:1986an}. A probable scenario was summarized by M.D.\ Scadron \emph{et al.} \cite{Scadron:2003ki} in 2002: \emph{``$\ldots$ The well-known $\Delta I=1/2$ rule $\Gamma(K_S\rightarrow \pi^+\pi^-)/\Gamma(K^+\rightarrow \pi^+\pi^0) \approx 450$ for nonleptonic weak $K_{2\pi}$ decays suggests that the parity-violating (PV) amplitude $\left<2\pi | H_w^{pv}|K_S\right>$ could be dominated by the $\Delta I=1/2$ weak transition $\left<\sigma| H_w^{pv}|K_S\right>$ $\ldots$''}. Although supported by other work \cite{Goswami:1972rn,Shifman:1976ge,Shabalin:1987jf} also this conjecture lacks yet to explain $S$-$P$ mixing.}. 
Unfortunately the solution of both problems requires a sizable yet unexplained mechanism for $S$-$P$ mixing \cite{Goswami:1972rn}. In the QLL$\sigma$M it is tempting to construct e.g.\ effective actions $S^{\,semi}_{e\!f\!f}[S\,\bar{q}q]$ or $S^{\,semi}_{e\!f\!f}[P\,\bar{q}q]$ for semi-strong $S \bar{q}q$- or $P \bar{q}q$-penguins \footnote{Dynamical generation yields the following semi-strong ``penguin'' effective actions in the local limit:
\begin{eqnarray} \lefteqn{S^{\,semi}_{e\!f\!f}[S\,\bar{q}q] = \int d^4z \;\,  G_F \; s_s \; e^{i\,\alpha} \, \frac{1}{N_c}\; \times} \nonumber \\[1mm]
 & \times & \Bigg( \overline{q^c_+}(z)\; \left( \begin{array}{cc} 0_3 & V_{{}_{CKM}} \\ 0_3 & 0_3 \\ \end{array}\right) \; [ \; \partial_\mu \left[ \frac{f\,S}{M} \right](z)\, , \, M_q \, ] \; \gamma^\mu \, P_{{}_L} \; \left( \begin{array}{cc} 0_3 & 0_3 \\ \overline{V}_{{}_{CKM}} & 0_3 \\ \end{array}\right)\; q_-(z) \nonumber \\
 &  & + \,\overline{q^c_+}(z)\; \left( \begin{array}{cc} 0_3 & 0_3 \\ \overline{V}_{{}_{CKM}} & 0_3 \\ \end{array}\right) \; [ \; \partial_\mu \left[  \frac{f\,S}{M} \right](z) \, , \, M_q \, ] \; \gamma^\mu \, P_{{}_L} \; \left( \begin{array}{cc} 0_3 & V_{{}_{CKM}} \\ 0_3 & 0_3 \\ \end{array}\right)
\; q_-(z) \Bigg) + \ldots \; , \\
\lefteqn{S^{\,semi}_{e\!f\!f}[P\,\bar{q}q] = \int d^4z \;\,  G_F \; s_p \; e^{i\,\alpha} \, \frac{2 i}{N_c}\; \times} \nonumber \\[1mm]
 & \times & \Bigg( \overline{q^c_+}(z)\; \left( \begin{array}{cc} 0_3 & V_{{}_{CKM}} \\ 0_3 & 0_3 \\ \end{array}\right) \; \partial_\mu \left[ f\,P \right](z) \; \gamma^\mu \, P_{{}_L} \; \left( \begin{array}{cc} 0_3 & 0_3 \\ \overline{V}_{{}_{CKM}} & 0_3 \\ \end{array}\right)\; q_-(z) \nonumber \\
 &  & + \,\overline{q^c_+}(z)\; \left( \begin{array}{cc} 0_3 & 0_3 \\ \overline{V}_{{}_{CKM}} & 0_3 \\ \end{array}\right) \; \partial_\mu \left[ f\,P \right](z) \; \gamma^\mu \, P_{{}_L} \; \left( \begin{array}{cc} 0_3 & V_{{}_{CKM}} \\ 0_3 & 0_3 \\ \end{array}\right) 
\; q_-(z) \Bigg) + \ldots \, ,
\end{eqnarray}
with $\left[ \frac{f\,S}{M} \right] \equiv (-\, 4\,i\, N_c \, |g|) \int \frac{d^4p}{(2\pi)^4} \,  \frac{1}{p^2- M_q^2 }  \, S \, \frac{1}{p^2- M_q^2}$ and $\left[ f\,P \right] \equiv (-\, 4\,i\, N_c \, |g|) \int \frac{d^4p}{(2\pi)^4} \,  \frac{1}{p^2- M_q^2 } \; \frac{1}{2} \{ P , \, M_q \} \, \frac{1}{p^2- M_q^2}$.} in the hope that they contribute to explain $S$-$P$ mixing. Unfortunately it turns out that relevant meson tadpoles constructed from these actions vanish, while non-tadpole graphs constructed from these actions don't contribute to the solution of the problems under consideration\footnote{$S^{\,semi}_{e\!f\!f}[S\,\bar{q}q]$ does not contribute to the cascade $K_S\rightarrow \sigma \rightarrow \pi^+\pi^-$, as PCVC enforces $S^{\,semi}_{e\!f\!f}[S\,\bar{q}q]=0$ for the ``flavour-neutral'' $\sigma$-meson, while $S^{\,semi}_{e\!f\!f}[P\,\bar{q}q]$ does not lead to an I=3/2 final state in $K^+\rightarrow \pi^+\pi^0$.}. We conclude that explanation of $S$-$P$ mixing \cite{Goswami:1972rn,Shifman:1976ge,Shabalin:1987jf} being probably of similar nature as $\pi^0$-$\eta$ or $\eta$-$\eta^\prime$ mixing \cite{Kleefeld:2005qs} requires further research.

{\bf Acknowledgments.} This work has been supported by the
FCT of the {\em Minist\'{e}rio da Ci\^{e}ncia, Tecnologia e Ensino Superior} \/of Portugal, under Grants no.\ SFRH/BDP/9480/2002, POCTI/\-FNU/\-49555/\-2002, and POCTI/FP/FNU/50328/2003.






\end{document}